\documentclass[superscriptaddress,twocolumn,prl]{revtex4}
\usepackage{graphicx}
\usepackage{amsmath,bm}
\usepackage{color}
\usepackage{hyperref}
\usepackage{amssymb,amsfonts}
\renewcommand{\d}{\mathrm{d}}
\begin{document}

\title{Mode-coupling theory for the dynamic heterogeneity in an aging glass: How Do Glassy Domains Grow?}
\author{Saroj Kumar Nandi}
\email{snandi@physics.iisc.ernet.in}
\affiliation{Centre for Condensed Matter Theory, Department of Physics, Indian
Institute of Science, Bangalore-560 012, India.}
\author{Sriram Ramaswamy}
\email{sriram@tifrh.res.in}
\altaffiliation{On leave at TIFR Centre for Interdisciplinary Sciences, 21 Brundavan Colony, Narsingi, Hyderabad 500 075, India.}
\affiliation{Centre for Condensed Matter Theory, Department of Physics, Indian Institute of Science, Bangalore-560 012, India.}
\begin{abstract}
We construct the equations for the growth kinetics of an aging structural glass within
mode-coupling theory through a non-stationary variant of the 3-density
correlator defined in Phys. Rev. Lett. {\bf 97}, 195701 (2006). We solve a
schematic form of the resulting equations to obtain the coarsening of
the dynamic heterogeneity, characterized via the 3-point correlator $\chi_3(t,t_w)$, as a function of waiting time $t_w$.
For a quench into the glass, we find that $\chi_3$ attains a peak value $\sim
t_w^{0.5}$ at $t -t_w \sim t_w^{0.8}$, providing a theoretical basis for the
numerical observations of Parisi [J. Phys. Chem. B \textbf{103}, 4128 (1999)]
and Kob and Barrat [Phys. Rev. Lett. \textbf{78}, 4581 (1997)]. The aging is not
``simple'': the $t_w$ dependence cannot be attributed to an evolving effective
temperature.
\end{abstract}
\maketitle

When a system is quenched below an ordering transition, domains of the
ordered phase appear and begin to grow \cite{bray94}, with
characteristic size given by the decay length of equal-time correlations
of the order parameter. The corresponding issue for the glass transition has
been examined numerically \cite{parisi99,parsaeian08,kob97}, using
susceptibilities and correlation lengths that capture the onset of amorphous
freezing \cite{chandan91,parisi99,berthier05,berthier_biroli_rmp2011}, 
but a quantitative theory of these observations has been lacking \cite{note1}.
Length-scale information similar to that
obtained from the 4-density correlator and related overlap functions
\cite{chandan91,parisi99,berthier05} has been shown \cite{biroli06} to be
contained in a certain three-point correlator $\chi_3(t)$, whose peak value,
and the time at which the peak
is attained, diverge \cite{biroli06} upon approaching the mode-coupling glass
transition \cite{spdas04}.

\begin{figure}
\includegraphics[width=7.6cm]{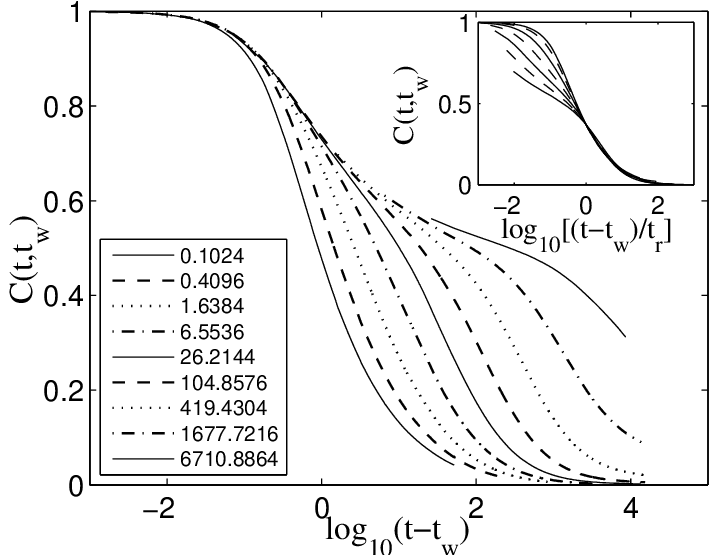}
\caption{The aging of the two-point function. The correlation function $C(t,t_w)$ as a function of $t$, for various waiting times $t_w$ shown in the legend. The decay with $t$ becomes progressively slower with
increasing $t_w$. The final parameter values are $T=1.0$ and $\lambda=2.0$.
{\bf Inset:} Scaling $t-t_w$ by $t_r$ yields a data collapse in the
$\alpha$-relaxation regime. Such ``simple aging'', however, is not seen in the
three-point function, Fig. \ref{chi4peakfit}.}
\label{corrfn}
\end{figure}

\begin{figure}
\includegraphics[width=6.6cm]{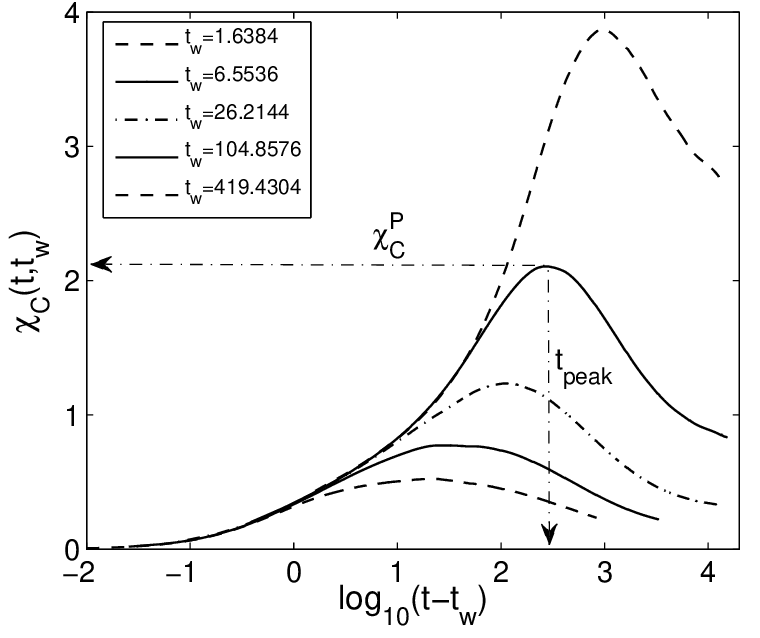}
\caption{The growth kinetics of glassy correlations. The three-point correlation function in an aging structural
glassy system, from schematic mode-coupling theory. The peak
value of $\chi_C(t,t_w)$ grows and shifts to
higher $t_{peak}$ with increasing waiting time $t_w$.}
\label{largelambda}
\end{figure}

\begin{figure}
\includegraphics[width=6.6cm]{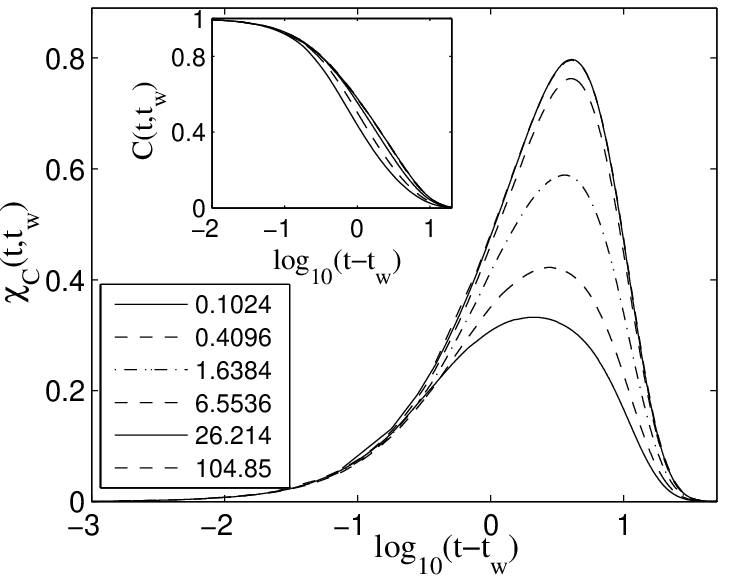}
\caption{Evolution of three-point function following a
quench to a point close to but on the liquid side of the
transition. Growth to saturation of $\chi_C(t,t_w)$ as a function of $t$ for various $t_w$ for
$\lambda=0.75$, corresponding to a quench into the liquid. {\bf Inset:} The
two-point correlator, for the same parameter values, shows a relaxation time
that grows progressively with increasing $t_w$ but approaches a finite value.}
\label{lowlambda}
\end{figure}

\begin{figure}
\includegraphics[width=7.6cm]{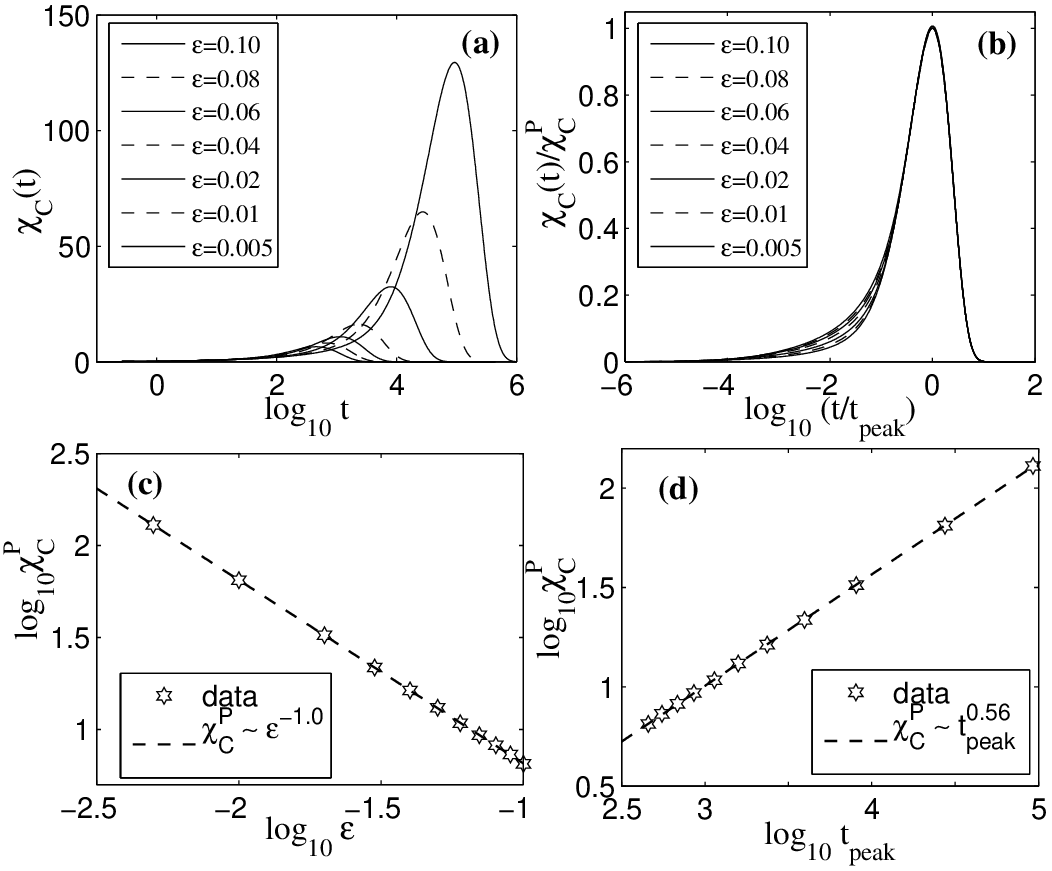}
\caption{Three-point function and correlation volume
in equilibrium. Three-point function and correlation volume in
equilibrium. (a) $\chi_C(t)$ as a function of $t$ for various
$\epsilon\equiv|\lambda-\lambda_c|$. (b) When $\chi_C(t)$ is scaled with
$\chi_C^P$ and time with $t_{peak}$, we obtain data collapse for large $t$. (c)
$\chi_C^P \sim \epsilon^{-1.0}$.
(d) $\chi_C^P\sim t_{peak}^{0.56}$}.
\label{ss_chi3scaling}
\end{figure}

\begin{figure}
\includegraphics[width=8.6cm]{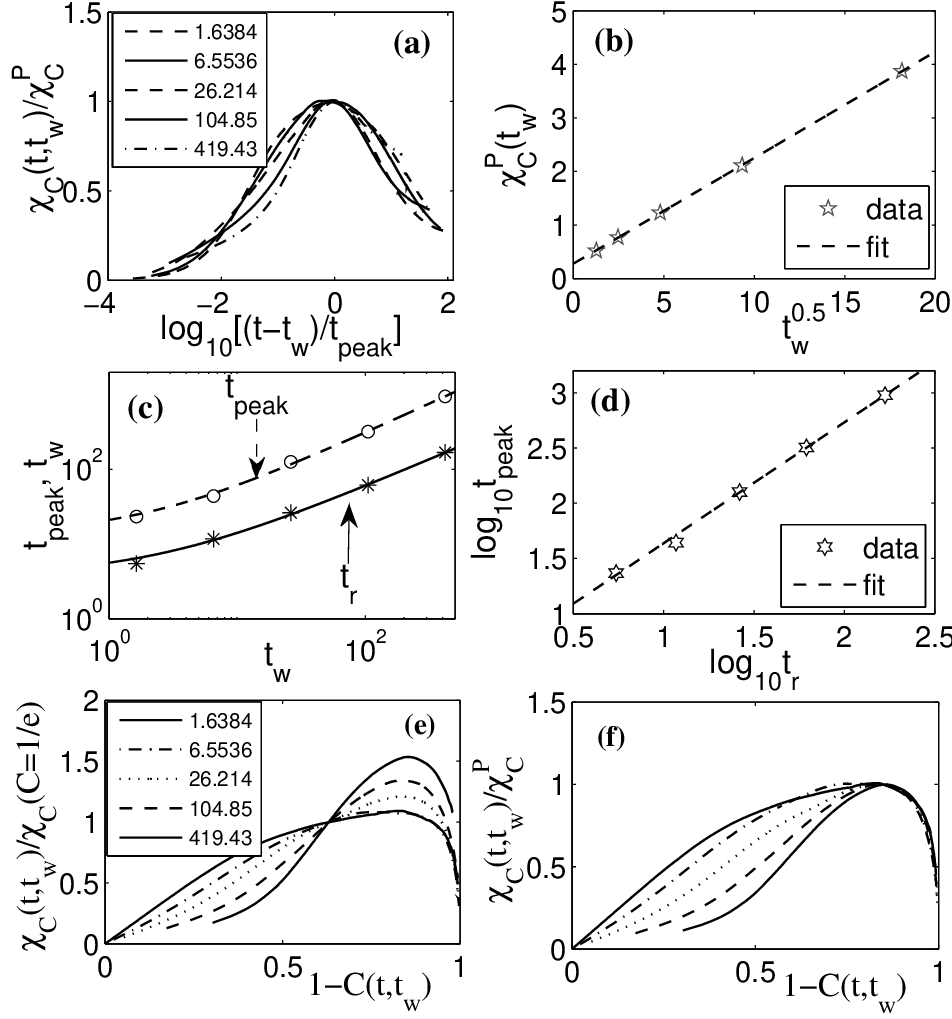}
\caption{Scaling of order-parameter correlations and
relaxation time following a quench into the glass: no
simple aging. (a) If we scale $\chi_C(t,t_w)$ by $\chi_C^P(t_w)$
and time by $t_{peak}$, the three-point function does not follow a master curve:
the behaviour at various $t_w$ is not the equilibrium dynamics at an evolving
effective temperature. This behaviour of $\chi_C$ belies the
expectation of ``simple aging'' suggested by the two-point function in Fig.
\ref{corrfn}. (b) The peak height $\chi_C^P(t_w) \propto t_w^{1/2}$. (c)
$t_{\text{peak}}$ and $t_r$ both grow as $t_w^{0.8}$. (d) $t_{\text{peak}}$ is
proportional to $t_r$. (e) If we scale $\chi_C(t,t_w)$ by
$\chi_C(C=1/e)$ and plot it as a function of $1-C(t,t_w)$, as in Parsaeian
{\it et al.} \cite{parsaeian08}, no data collapse is found. $t_w$ values are
shown in the legend. (f) Scaling $\chi_C(t,t_w)$ by
$\chi_C^P$ and plotting it as a function of $1-C(t,t_w)$ shows data collapse in
the regime of $\alpha$-relaxation. Waiting times as in (e).}
\label{chi4peakfit}
\end{figure}

In this work we present a theory of \textit{the coarsening of glassy
order}, using a non-stationary generalization $\chi_3(t,t_w)$,
whose peak value $\Omega(t_w)$ is the correlation volume as a
function of the waiting time $t_w$ since the quench. We formulate our
calculation in the framework of the fluctuating hydrodynamics of a dense
liquid, and obtain results using mode-coupling theory (MCT)
\cite{goetzebook,spdas04,reichman05}, in a schematic approach
\cite{franz95}.
Figures \ref{corrfn} - \ref{chi4peakfit} summarize our results. We find that
$\Omega(t_w)$ grows without bound for a quench into the MCT glass (Fig.
\ref{largelambda}), as $t_w^{0.5}$, and the relaxation time as $t_w^{0.8}$ (Fig.
\ref{chi4peakfit}), in agreement respectively with the computer
experiments of Parisi \cite{parisi99} 
and Kob and Barrat \cite{kob97}. 
As effects beyond MCT cut off the transition, the coarsening in experiments, simulations, or a complete theory will cease at long enough times, but typical simulations do not explore these asymptotically long time scales and can therefore be compared usefully to our MCT coarsening predictions.
The three-point function, Fig. \ref{chi4peakfit}, shows features incompatible
with ``simple aging'' \cite{kob97,sunil09} 
but qualitatively similar to \cite{parsaeian08}. 
For a quench to a distance $\epsilon$ from the threshold value on the \textit{liquid}
side, $\Omega$ grows to saturation (Fig. \ref{lowlambda}), reaching an
equilibrium value
$\sim \epsilon^{-1}$, with a relaxation time $\sim \epsilon^{-1.8}$ (Figure~\ref{ss_chi3scaling}).

MCT is an analytically tractable approximation to equilibrium
liquid-state dynamics that yields a glass transition in a homogeneous system,
which is why it is so widely used despite its shortcomings
\cite{reichman05,berthier_biroli_rmp2011}. In
order to extend MCT to describe non-stationary states such as coarsening we work
with a general field-theoretic approach \cite{bouchaud96,reichman05}, taking
care not to use results like the Kubo formulae and fluctuation-dissipation
relations (FDR), which are justified only in equilibrium treatments
\cite{zaccarelli02,kawasaki03}.
We start with the equations of fluctuating hydrodynamics for the velocity and
density fields for an isothermal compressible fluid, extended to large
wavenumbers so as to take into account the modes around the structure factor
peak \cite{spdas04}. In order to obtain an equation for the density field alone,
we eliminate the velocity while retaining momentum conservation but ignoring
inertia. This yields the dynamical equation
\begin{align}
\label{kdep1}
\frac{\partial \delta \rho_k(t)}{\partial t} +K_1\delta\rho_k(t)=\frac{K_2}{2}\int_{\bf q}\mathcal{V}_{k,q}\delta\rho_q(t)\delta\rho_{k-q}(t)+f_k(t),
\end{align}
for the Fourier-transformed density fluctuation $\delta\rho_k(t)$ at
wavevector $\mathbf{k}$, with
$\mathcal{V}_{k,q}=\mathbf{k}\cdot[\mathbf{q}c_q+(\mathbf{k}-\mathbf{q})c_{k-q}]
$, $K_1={k_BT}/{S_kD_L}$ and $K_2={k_BT}/{D_Lk^2}$.
Eq. (\ref{kdep1}) can be viewed as the no-inertia limit
of Eq. (4.1) of \cite{kawasaki03}. Here $D_L=(\zeta+4\eta/3)/\rho_0$ is the
longitudinal damping, where $\zeta$ and $\eta$ are the bare shear and bulk
viscosities, $k_B T$ is Boltzmann's constant times temperature, $S_k$ and $c_k$
are the equilibrium static structure factor and direct correlation function
respectively and the noise $f_k(t)$ obeys 
\begin{equation}
\label{kdep2}
\langle f_k(t)f_{k'}(t')\rangle=\frac{2k_BT}{D_L}\rho_k(t)\delta(k+k')\delta(t-t').
\end{equation}
From diagrammatic perturbation theory \cite{bouchaud96,reichman05}, we
construct the equations of motion for the correlation function
$C_k(t,t_w)=\langle \delta\rho_k(t)\delta\rho_{-k}(t_w)\rangle$ and response
function $R_k(t,t_w)=\langle \partial \delta\rho_k(t)/\partial
f_{-k}(t_w)\rangle$:
\begin{subequations}
\label{expliciteq}
\begin{align}
\label{expliciteqa}
\frac{\partial C_k(t,t_w)}{\partial t} =& -K_1C_k(t,t_w)+\int_0^{t_w}\d s
D_k(t,s)R_k(t_w,s) \nonumber\\
 &+\int_{0}^t \d s \Sigma_k(t,s)C_k(s,t_w),\\
\label{expliciteqb}
\frac{\partial R_k(t,t_w)}{\partial t} =& \delta(t-t_w)-K_1R_k(t,t_w) \nonumber \\
&+\int_{t_w}^t \d s \Sigma_k(t,s)R_k(s,t_w),   
\end{align}
\end{subequations}
with $D_k(t,t') = ({2k_BT}/{D_L})\rho_k(t)\delta(t-t') + M_k(t,t')$, $M_k(t,t')=({K_2^2}/{2}) \int_{\bf
q}\mathcal{V}_{k,q}^2C_q(t,t')C_{k-q}(t,t')$ and $\Sigma_k(t,t')=K_2^2\int_{\bf q} \mathcal{V}_{k,q}^2R_q(t,t')C_{k-q}(t,t')$. The contribution
to (\ref{expliciteqa}) from the first term in $D_k(t,t')$ vanishes due to
causality. Franz and Hertz \cite{franz95} obtained schematic equations similar
to (\ref{expliciteqa}) and (\ref{expliciteqb}) for the Amit-Roginsky model
\cite{amit79}.

How are the input quantities $K_1$ and $\mathcal{V}_{k,q}$ in
equations (\ref{expliciteqa}) and (\ref{expliciteqb}) defined for the case of a
quench? A comparison with the treatment of Zaccarelli \textit{et al.}
\cite{zaccarelli02} is useful here. The $\mathcal{V}_{k,q}$ term in
(\ref{kdep1}) and (\ref{expliciteq}) involves the ``residual interactions'' in
\cite{zaccarelli02}. We define our quench to be an abrupt increase in the
interaction strength, implying that $\mathcal{V}_{k,q}$ should be evaluated at
the final parameter value. To determine $K_1$, which must now be a
time-dependent quantity as we are dealing with a non-stationary state, we
insist, as in \cite{cugliandolo93}, that for $\tau=(t-t_w) \ll t_w$ Eq.
(\ref{expliciteq}) obeys time-translation invariance and the FDR. This leads,
after some algebra, to 
\begin{align}
\label{k1oft}
K_1(t)S_k=&TR_k(0)+K_2^2\int_0^t \int_{\bf q} \mathcal{V}_{k,q}^2C_{k-q}(t,s) \times \nonumber\\
&\big[\frac{1}{2}C_q(t,s)R_k(t,s)+R_q(t,s)C_k(t,s)\big]\d s.
\end{align}
In \cite{zaccarelli02} the term corresponding to $K_1$ enters
through the equal-time density correlator. The latter being time-dependent
in a coarsening situation, it is natural that $K_1$ should change in time.

To obtain the equation of motion
for the growth kinetics of glassy correlations, we look at the behaviour of
our nonstationary generalization of the three-density
correlation function \cite{biroli06} mentioned above. We introduce in the
free-energy functional a one-body term $\epsilon(\mathbf{r}) \rho(\mathbf{r})$,
coupling the density to an external potential $\epsilon(\mathbf{r})$ and leading
on average to an inhomogeneous shift $\delta m(\mathbf{r})$ in the mean density
field \cite{sarojunpub}. We work in the limit where $\epsilon(\mathbf{r})$
and hence $\delta m(\mathbf{r})$ are uniform, so that the Fourier
transform $\delta m_{\mathbf{k}}$ has non-zero weight $\delta m_0$ only for
wavevector $\mathbf{k}=0$. We will see that this suffices for the purpose of
extracting the correlation volume. The resulting generalized Langevin equation
for
$\rho$, to first order in the background density $\delta m_0$ which encodes the
effects of the field, is \cite{saroj11,note2}  
\begin{align}
\label{withpotential}
\frac{\partial \delta \rho_k(t)}{\partial t} &+ K_1(t)\delta\rho_k(t)-\frac{k_BTc_k\delta m_0}{D_L}\delta\rho_{k}(t) \nonumber\\
=&\frac{K_2}{2}\int_{\bf
q}\mathcal{V}_{k,q}\delta\rho_q(t)\delta\rho_{k-q}(t)+f_k(t).
\end{align}
Let $\tilde{C}_k(t,t_w)$ and $\tilde{R}_k(t,t_w)$ denote the 
$\delta m_0$-dependent two-point correlation and response functions
implied by (\ref{withpotential}). As we are working in a non-stationary state,
we must define separate 3-point quantities analogous to $\chi_3$ in
\cite{biroli06} for $\tilde{C}_k$ and $\tilde{R}_k$: $\chi^C_k(t,t_w) =
{\partial \tilde{C}_k(t,t_w)}/{\partial \delta m_0}|_{\delta m_0\to 0}$ and
$\chi^R_k(t,t_w)= {\partial \tilde{R}_k(t,t_w)}/{\partial \delta m_0}|_{\delta
m_0\to0}$, with equations of motion 
\begin{subequations}
\label{chiRC}
\begin{align}
 \label{chiRCa}
\frac{\partial \chi_k^{R}(t,t_w)}{\partial t}& + K_1(t)\chi_k^{R}(t,t_w) 
=\int_{t_w}^t \d s {\Sigma}_k(t,s)\chi^{R}_k(s,t_w) \nonumber\\
&+\int_{t_w}^t \d s \tilde{\Sigma}'_k(t,s){R}_k(s,t_w)
+\mathcal{S}^R_k(t,t_w), \\
\label{chiRCb}
\frac{\partial \chi^{C}_k(t,t_w)}{\partial t}& +K_1(t)\chi^{C}_k(t,t_w) 
=\int_0^{t_w}\d s {M}_k(t,s)\chi^{R}_k(t_w,s) \nonumber\\
\hspace{-1cm}+\int_0^{t_w}\d s & \tilde{M}'_k(t,s){R}_k(t_w,s) +\int_{0}^t \d s {\Sigma}_k(t,s)\chi^{C}_k(s,t_w)\nonumber\\
+&\int_{0}^t \d s \tilde{\Sigma}'_k(t,s){C}_k(s,t_w) +\mathcal{S}^C_k(t,t_w),
\end{align}
\end{subequations}
where $\tilde{\Sigma}'_k(t,s)=\partial \tilde{\Sigma}_k(t,s)/\partial\delta
m_0|_{\delta m_0\to0}$,  and $\tilde{M}'_k(t,s)={\partial
\tilde{M}_k(t,s)}/{\partial\delta m_0}|_{\delta m_0\to0}$, $\tilde{M}_k(t,s)$
and $\tilde{\Sigma}_k(t,s)$ are quantities corresponding to $M$ and $\Sigma$ but
evaluated in the presence of $\epsilon(\mathbf{r})$.
The expressions for the source terms $\mathcal{S}^R_k(t,t_w)$ and
$\mathcal{S}^C_k(t,t_w)$ are given in the supplementary information (SI) and the straightforward but tedious derivation of (\ref{chiRCa}) and (\ref{chiRCb}) will be presented in a subsequent paper \cite{sarojunpub}.

Simplified integral equations keeping track of time-dependence alone
\cite{leutheusser84,kirkpatrick85,brader09} have proved invaluable in
extracting meaningful results from MCT within a manageable calculation. In this
spirit, we suppress dependence on wavevector $\mathbf{k}$ and write the
self-energies in (\ref{expliciteq}) and (\ref{chiRC}) as $M(t,s)=2\lambda
C^2(t,s)$ and $\Sigma(t,s)=4\lambda R(t,s)C(t,s)$, yielding equations for the
two- and three-point correlation and response functions which we now
denote $C(t,t_w)$, $R(t,t_w)$, $\chi_C(t,t_w)$ and $\chi_R(t,t_w)$.
We solve the resulting schematic versions of
(\ref{expliciteq}), (\ref{k1oft}) and (\ref{chiRC}), whose detailed forms are given in the SI section, using the algorithm developed by Kim and Latz
\cite{kim01,herzbach00}, to give the results quoted at the start of this paper.
First, aging can clearly be seen in the behaviour of $C(t,t_w)$ in Fig.
\ref{corrfn}. Second and more important is the characteristic non-monotone
behavior of $\chi_C(t,t_w)$, and its dependence on $t_w$ and interaction
strength $\lambda$ (Fig. \ref{largelambda}). For a fixed initial
condition corresponding in our schematic
approach to a liquid with negligible correlations, we examine in particular how
$\chi_C(t,t_w)$ as a function of $t$ changes with $t_w$, for values of $\lambda$
corresponding to the liquid and the glass phase. Recall that $\lambda$ defines
the point to which the system is quenched. For $\lambda$ in the liquid phase but
close to the transition we find, as expected, that $\chi_C(t,t_w)$ attains a
peak value $\chi_C^P$ at a time $t_{peak}$, with both $\chi_C^P$ and $t_{peak}$ 
growing with $t_w$ but saturating to finite values as shown in Fig. \ref{lowlambda}.
The final peak value of $\chi_C(t,t_w)$ grows as $(\lambda-\lambda_c)^{-1}$ and
$\chi_C^P\sim t_{peak}^{0.56}$ (Fig. \ref{ss_chi3scaling}). These final values, obtained at
$t_w\to\infty$ are the equilibrium values of the corresponding quantities.
In the notation of \cite{biroli06} we are working at $\mathbf{q}_0\to 0$ and our
results are consistent with theirs in that limit. A more detailed comparison
with \cite{biroli06} or \cite{karmakar09}, including an estimate of the
correlation \textit{length} requires a calculation of the sensitivity of
two-point functions to a spatially varying potential.

For a quench into the glassy region, $\lambda = 2.0$, as shown in
Fig. \ref{largelambda}, $\chi_C^P$ grows without bound. In more detail (Fig.
\ref{chi4peakfit}), $\chi_C^P \sim t_w^a$ at a time $t_{peak} \sim t_w^b$, with
increasing $t_w$. The exponents $a \simeq 0.5$ in agreement with
simulations \cite{parisi99} and $b \simeq 0.8$.
From the two-point correlator $C(t,t_w)$ we find a relaxation time $t_r \simeq
t_{peak}/4$ close to the transition. Thus our result implies $t_r\sim
t_w^{0.8}$, in agreement with
the numerical experiment of Ref. \cite{kob97}; the relation between $t_r$
and $t_{peak}$ remains to be tested. $\chi_C^P$ measures an effective
correlation volume, so that its growth is the claimed coarsening of glassy
structure, and is consistent with the idea of a growing ``domain size''.
Regardless of the precise values obtained, it is significant that our theory and
the simulations of \cite{parisi99,parsaeian08,kob97}
all find a total structure factor $\chi_C^P$ growing very sublinearly in time. Our scaling laws differ quantitatively (Fig. \ref{chi4peakfit}e) from those of \cite{parsaeian08},
perhaps because we measure different quantities.
However, if we scale $\chi_C(t,t_w)$ by $\chi_C^P$ and plot them as
a function of $1-C(t,t_w)$, data collapse is obtained in the $\alpha$-relaxation
regime (Fig. \ref{chi4peakfit}f). We do not claim to understand
the origin of this scaling or, for that matter, that of Ref.
\cite{parsaeian08}. A similar calculation \cite{sarojunpub} for the three-point
correlation function for a $p$-spin spin{\color{magenta}-}glass model with $p=3$
finds again a growing $\chi_C^P$, but slower than for the present problem.

We emphasize that the $t_w$-dependent properties we extract do not
correspond to those of an equilibrium system at an evolving $\lambda$ or 
temperature. Had it been so, scaling $\chi_C(t,t_w)$ by
$\chi_C^P(t_w)$ and time by $t_{peak}$ would have given data collapse for all
$t_w$ as for the equilibrium case (Fig. \ref{ss_chi3scaling}).
Fig. \ref{chi4peakfit} shows the absence of such collapse even for larger $t_w$.
It would appear that the 3-point correlator is more sensitive to departures from
``simple aging'', and an interpretation in terms of an evolving effective
temperature, than the two-time correlation function \cite{kob97,sunil09} (see
inset of Fig. \ref{corrfn}). Perhaps the monotone decay of the latter masks such
deviations or, more likely, $\chi_C$ carries additional, independent
information.

We close by summarising the achievements of this work. We have
shown that mode-coupling theory adapted to describe non-stationary states
captures the key features of the emergence and coarsening of glassy order from a
liquid. Through the evolution of a three-point function we have shown that the
glassy correlation volume grows as $t_w^{0.5}$ with waiting time $t_w$, slower
than domain volumes in conventional coarsening, and the relaxation time of the
glass grows as $t_w^{0.8}$. These theoretical growth laws are supported by
simulation studies \cite{parisi99,kob97}, and the broad features
we observe are similar to those in \cite{parsaeian08}. In an experimental
realization, if the quench is below the MCT transition but above a putative
ideal glass transition at, say, the Kauzmann temperature $T_K$, activated
processes \cite{sarika08} outside the scope of MCT should cut off the growth.
Presumably a quench below $T_K$ will give indefinite growth of a different
glassy length scale
\cite{berthier_kob_2011,cammarota_biroli_2011,kurchan11} with a
form not predicted by MCT. In results to be presented separately
\cite{sarojunpub} we find further that an imposed shear-rate $\dot{\gamma}$ cuts
off aging and coarsening at $t_w \sim 1/\dot{\gamma}$ in the glassy region and
$t_w = \min(t_r,1/\dot{\gamma})$ in the fluid. {Since the relaxation time goes
as $t_w^{0.8}$, this should imply that $t_r$ or $t_{peak}$ should vary as
$\dot{\gamma}^{-0.8}$}. We look forward to experimental tests of our results.

\begin{acknowledgments}
We thank C. Dasgupta for valuable suggestions and a critical reading of our
paper and N. Menon for enlightening comments. We also thank S.M. Bhattacharyya,
B. Kim, K. Miyazaki, S. Sastry, D. Sen, S.P. Singh and E. Zaccarelli
for discussions. SKN was supported in part by the University Grants
Commission and SR by a J.C. Bose Fellowship from the Department of
Science and Technology. 
\end{acknowledgments} 

\vspace{1cm}
\begin{center}
{\bf \underline{Supplementary information (SI)}}
\end{center}
\section*{The schematic form of the equations and the source terms}
The schematic form of the equations (3a) and (3b), is obtained by following the outline given in the paper. The final forms will be:
\begin{align}
\frac{\partial C(t,t_w)}{\partial t} =& -\mu(t) C(t,t_w)+ 2\lambda \int_0^{t_w}C^2(t,s)R(t_w,s) \d s \nonumber \\
&+4\lambda\int_0^t C(t,s)R(t,s)C(s,t_w)\d s, \nonumber \\
\frac{\partial R(t,t_w)}{\partial t} =& \delta(t-t_w)-\mu(t) R(t,t_w) \nonumber \\
&+4\lambda\int_{t_w}^t R(t,s)C(t,s)R(s,t_w)\d s
\end{align}
where $\mu(t)$ is the schematic version of $K_1(t)$:
\begin{equation}
\mu(t)=T+6\lambda\int_0^t C^2(t,s)R(t,s)\d s.
\end{equation}

Eqs. (7a) and (7b) contain two ``source'' terms which we present in detail here.
To proceed, let us define
\begin{align}
\omega_k(t)=&\frac{K_2^2}{S_k}\int_0^t \int_{\bf q} \mathcal{V}_{k,q}^2\bigg[\chi^C_{k-q}(t,s)\bigg\{\frac{1}{2}C_q(t,s)R_k(t,s)\nonumber\\
&+R_q(t,s)C_k(t,s)\bigg\}\d s \nonumber\\
+&C_{k-q}(t,s)\bigg\{\frac{1}{2}\chi^C_q(t,s)R_k(t,s)+\frac{1}{2}C_q(t,s)\chi^R_k(t,s)\nonumber\\
&+\chi^R_q(t,s)C_k(t,s)+R_q(t,s)\chi_k^C(t,s)\bigg\}\bigg]\d s.
\end{align}
Then the source terms can be written in the form
\begin{align}
\mathcal{S}_k^R(t,t_w)=\frac{k_BTc_k}{D_L}R_k(t,t_w)-\omega_k(t)R_k(t,t_w),
\nonumber\\
\mathcal{S}_k^C(t,t_w)=\frac{k_BTc_k}{D_L}C_k(t,t_w)-\omega_k(t)C_k(t,t_w)
\end{align}
The equations for the three-point correlators are also schematicised in a similar way as stated in the paper. The final schematic forms of equations (7a) and (7b) will be
\begin{widetext}
\begin{align}
\label{chi3schematic}
\frac{\partial \chi_R(t,t_w)}{\partial t} &+\mu(t)\chi_R(t,t_w)=4\lambda\int_{t_w}^t R(t,s)C(t,s)\chi_R(s,t_w)\d s \nonumber +4\lambda\int_{t_w}^t R(t,s)\chi_C(t,s)R(s,t_w)\d s \nonumber\\
+&4\lambda\int_{t_w}^t \chi_R(t,s)C(t,s)R(s,t_w)\d s +\mathcal{S}_R(t,t_w)\\
\frac{\partial \chi_C(t,t_w)}{\partial t} & +\mu(t)\chi_C(t,t_w)= 4\lambda\int_0^{t_w}C(t,s)\chi_C(t,s)R(t_w,s)\d s +2\lambda\int_0^{t_w} C^2(t,s)\chi_R(t_w,s)\d s \nonumber\\
+&4\lambda\int_0^{t} C(t,s)R(t,s)\chi_C(s,t_w)\d s +4\lambda\int_0^{t} \chi_C(t,s)R(t,s)C(s,t_w)\d s \nonumber\\
+&4\lambda\int_0^{t} C(t,s)\chi_R(t,s)C(s,t_w)\d s +\mathcal{S}_C(t,t_w)
\end{align}
\end{widetext}
with the source terms given as $\mathcal{S}_R(t,t_w)=[1-\omega(t)]R(t,t_w)$ and $\mathcal{S}_C(t,t_w)=[1-\omega(t)]C(t,t_w)$ where $\omega(t)$, the schematic form of $\omega_k(t)$, is given as
\begin{align}
\omega(t)&=12\lambda \int_0^t C(t,s)\chi_C(t,s)R(t,s)\d s \nonumber\\
&+6\lambda\int_0^t C^2(t,s)\chi_R(t,s)\d s. 
\end{align}


\begin{thebibliography}{35}
\bibitem[{not({\natexlab{a}})}]{note1}
\bibinfo{note}{The aging of the time-decay of two-point density correlations
  has been studied in experiments \cite{bonn02}, numerical simulations
  \cite{kob97}, mode-coupling theories \cite{latz00} and generalized Langevin
  treatments \cite{pedro10} on structural glasses, and in related theoretical
  work on spin-glasses \cite{cugliandolo93,kim01,sunil09}.}
\bibitem[{not({\natexlab{b}})}]{note2}
\bibinfo{note}{We introduce the external potential in the coarse-grained
  free-energy functional whereas \cite{biroli06} do so in the microscopic
  Hamiltonian, where it must affect all degrees of freedom, fast and slow.}
\end{thebibliography}
\end{document}